\documentclass[a4paper,
              ]{jacow}

\makeatletter
\ifboolexpr{bool{xetex}}                 
 {\renewcommand{\Gin@extensions}{.pdf,%
                    .png,.jpg,.bmp,.pict,.tif,.psd,.mac,.sga,.tga,.gif,%
                    .eps,.ps,%
                    }}{}
\makeatother

\ifboolexpr{bool{xetex} or bool{luatex}} 
 {}                                      
 {\usepackage[utf8]{inputenc}}           

\usepackage{todonotes}
\usepackage[USenglish]{babel}
\usepackage[acronym,nomain,nonumberlist,nopostdot]{glossaries}
\newacronym{cern}{CERN}{European Organization for Nuclear Research}
\newacronym{infn}{INFN}{Istituto Nazionale di Fisica Nucleare}
\newacronym{kit}{KIT}{Karlsruhe Institute of Technology}
\newacronym{desy}{DESY}{Deutsches Elektronen-Synchrotron}
\newacronym{ess}{ESS}{European Spallation Source}
\newacronym{peg}{PEG}{Polish Electronics Group}

\newacronym{slc}{SLC}{Stanford Linear Collider}
\newacronym{lep}{LEP}{Large Electron-Positron Collider}
\newacronym{lhc}{LHC}{Large Hadron Collider}
\newacronym{hllhc}{HL-LHC}{High-Luminosity Large Hadron Collider}
\newacronym{helhc}{HE-LHC}{High Energy Large Hadron Collider}
\newacronym{fcc}{FCC}{Future Circular Collider}

\newacronym{cms}{CMS}{Compact Muon Solenoid}
\newacronym{atlas}{ATLAS}{A Toroidal LHC ApparatuS}
\newacronym{alice}{ALICE}{A Large Ion Collider Experiment}
\newacronym{lhcb}{LHCb}{Large Hadron Collider beauty}
\newacronym{totem}{TOTEM}{TOTal Elastic and diffractive cross section Measurement}
\newacronym{moedal}{MoEDAL}{Monopole and Exotics Detector at the Large Hadron Collider}
\newacronym{lhcf}{LHCf}{Large Hadron Collider forward}
\newacronym{delphi}{DELPHI}{Detector with Lepton, Photon and Hadron Identification}
\newacronym{cdf}{CDF}{Collider Detector at Fermilab}

\newacronym{L1Trigger}{L1 trigger}{Level-1 trigger}
\newacronym{L2Trigger}{L2 trigger}{Level-2 trigger}
\newacronym{hlt}{HLT}{High-Level Trigger}
\newacronym{bu}{BU}{Builder Unit}
\newacronym{fu}{FU}{Filter Unit}
\newacronym{ecal}{ECAL}{Electromagnetic Calorimeter}
\newacronym{hcal}{HCAL}{Hadron Calorimeter}
\newacronym{am}{AM}{Associative Memory}
\newacronym{cam}{CAM}{Content-Addressable Memory}
\newacronym{dt}{DT}{Drift Tube}
\newacronym{csc}{CSC}{Cathode Strip Chamber}
\newacronym{ctc}{CTC}{Central Tracking Chamber}
\newacronym{rpc}{RPC}{Resistive Plate Chamber}
\newacronym{llrf}{LLRF}{Low-Level Radio Frequency}

\newacronym{xft}{XFT}{eXtremely Fast Tracker}
\newacronym{svt}{SVT}{Secondary Vertex Trigger}
\newacronym{svx}{SVXII}{Silicon VerteX detector for run II}
\newacronym{ftk}{FTK}{Fast TracKer}
\newacronym{rod}{ROD}{Read-Out Driver}

\newacronym{2s}{2S}{Strip-Strip}
\newacronym{ps}{PS}{Pixel-Strip}
\newacronym{dtc}{DTC}{Data, Trigger and Control Board}

\newacronym{cic}{CIC}{Concentrator Integrated Circuit}
\newacronym{gbt}{GBT}{GigaBit Transceiver}
\newacronym{cbc}{CBC}{CMS Binary Chip}
\newacronym{mpa}{MPA}{Macro-Pixel ASIC}
\newacronym{ssa}{SSA}{Strip Sensor ASIC}
\newacronym{do}{DO}{Data Organizer}

\newacronym{adc}{ADC}{Analog-to-Digital Converter}
\newacronym{dac}{DAC}{Digital-to-Analog Converter}
\newacronym{daq}{DAQ}{Data Acquisition}
\newacronym{spi}{SPI}{Serial Peripheral Interface bus}
\newacronym{i2c}{I2C}{Inter-Integrated Circuit}
\newacronym{gpio}{GPIO}{General-Purpose Input/Output}
\newacronym{pcie}{PCIe}{Peripheral Component Interconnect Express}
\newacronym{axi4}{AXI4}{Advanced eXtensible Interface version 4}
\newacronym{dma}{DMA}{Direct Memory Access}

\newacronym{amchip}{AM Chip}{Associative Memory Chip}
\newacronym{asic}{ASIC}{Application-Specific Integrated Circuit}
\newacronym{fpga}{FPGA}{Field-Programmable Gate Array}
\newacronym{ip}{IP}{Intellectual Property}
\newacronym{pla}{PLA}{Programmable Logic Array}
\newacronym[longplural={Random Access Memories}]{ram}{RAM}{Random Access Memory}
\newacronym{cpu}{CPU}{Central Processing Unit}
\newacronym{pcb}{PCB}{Printed Circuit Board}
\newacronym{qsfp}{QSFP+}{Quad Small Form-factor Pluggable}
\newacronym{sfp}{SFP}{Small Form-factor Pluggable}
\newacronym{rtm}{RTM}{Rear Transition Module}
\newacronym{amc}{AMC}{Advanced Mezzanine Carrier}
\newacronym{atca}{ATCA}{Advanced Telecommunications Computing Architecture}
\newacronym{mtca}{MTCA}{Micro Telecommunications Computing Architecture}
\newacronym{fmc}{FMC}{FPGA Mezzanine Card}
\newacronym{ddr2}{DDR2}{Double Data Rate version 2}
\newacronym{rldram}{RLDRAM}{Reduced Latency DRAM}
\newacronym[longplural={Dynamic Random-Access Memories}]{dram}{DRAM}{Dynamic Random-Access Memory}
\newacronym[longplural={Static Random-Access Memories}]{sram}{SRAM}{Static Random-Access Memory}
\newacronym{dsp}{DSP}{Digital Signal Processing}
\newacronym{lvds}{LVDS}{Low-Voltage Differential Signaling}
\newacronym{ipmc}{IPMC}{Intelligent Platform Management Controller}
\newacronym{ipmi}{IPMI}{Intelligent Platform Management Interface}

\newacronym{tsmc}{TSMC}{Taiwan Semiconductor Manufacturing Company}

\newacronym{prm}{PRM}{Pattern Recognition Mezzanine}

\newacronym{rf}{RF}{Radio Frequency}
\newacronym{lut}{LUT}{Lookup Table}
\newacronym{bram}{BRAM}{Block RAM}
\newacronym{fifo}{FIFO}{First-In, First-Out}
\newacronym{rtl}{RTL}{Register-Transfer Level}
\newacronym{stl}{C++ STL}{C++ Standard Template Library}
\newacronym{hdl}{HDL}{Hardware Description Language}
\newacronym{vhdl}{VHDL}{Very High Speed Integrated Circuit Hardware Description Language}
\newacronym{html}{HTML}{Hypertext Markup Language}
\newacronym{uml}{UML}{Unified Modeling Language}
\newacronym{cmssw}{CMSSW}{CMS Software}
\newacronym{edm}{EDM}{Event Data Model}
\newacronym{gui}{GUI}{Graphical User Interface}

\newacronym{dc}{DC}{Don't Care}
\newacronym{pca}{PCA}{Principal Component Analysis}
\newacronym{idb}{IDB}{Intelligent Data Buffer}
\newacronym{tmtt}{TMTT}{Time Multiplexing Track Trigger}
\newacronym{mp7}{MP7}{Imperial Master Processor Virtex-7}
\newacronym{ascii}{ASCII}{American Standard Code for Information Interchange}
\newacronym{jtag}{JTAG}{Joint Test Action Group}
\newacronym{hssl}{HSSL}{High-Speed Serial Link}

\newacronym{lan}{LAN}{Local Area Network}
\newacronym{udp}{UDP}{User Datagram Protocol}
\newacronym{tcp}{TCP}{Transmission Control Protocol}
\newacronym{ipnet}{IP}{Internet Protocol}

\newacronym{epics}{EPCIS}{Experimental Physics and Industrial Control System}

\ifboolexpr{bool{jacowbiblatex}}
 {%
  \addbibresource{jacow-test.bib}
  \addbibresource{biblatex-examples.bib}
 }{}

\newcommand\blfootnote[1]{%
  \begingroup
  \renewcommand\thefootnote{}\footnote{#1}%
  \addtocounter{footnote}{-1}%
  \endgroup
}

\begin{document}

\title{The \NoCaseChange{ESS} \NoCaseChange{FPGA} Framework and its Application on the \NoCaseChange{ESS} \NoCaseChange{LLRF} System}

\author{C. Amstutz, M. Donna, A. J. Johansson\textsuperscript{1}, European Spallation Source ERIC, Lund, Sweden\\
        M. Mohammednezhad, Sigma Connectivity AB, Lund, Sweden\\
	\textsuperscript{1}also at Lund University, Lund, Sweden}

\maketitle


\blfootnote{The authors of this work grant the arXiv.org and LLRF Workshop's International Organizing Committee a non-exclusive and irrevocable license to distribute the article, and certify that they have the right to grant this license.}


\begin{abstract}
  The functions of the \gls{llrf} system at \gls{ess} are implemented on different \gls{fpga} boards in a \gls{mtca} crate.
  Besides the algorithm, code that provides access to the peripherals connected to the \gls{fpga} is necessary.
  In order to provide a common platform for the \gls{fpga} developments at \gls{ess}---the \gls{ess} \gls{fpga} Framework has been designed.
  The framework facilitates the integration of different algorithms on different \gls{fpga} boards.
  Three functions are provided by the framework:
  (1) Communication interfaces to peripherals, e.g.\ \glspl{adc} and on-board memory,
  (2) Upstream communication with the control system over \gls{pcie}, and
  (3) Configuration of the on-board peripherals.
  To keep the framework easily extensible by \gls{ip} blocks and to enable seamless integration with the Xilinx design tools, the \gls{axi4} bus is the chosen communication interconnect.
  Furthermore, scripts automatize the building of the \gls{fpga} configuration, software components and the documentation.
  The \gls{llrf} control algorithms have been successfully integrated into the framework.
\end{abstract}


\section{Introduction}

By 2023, \gls{ess} will be the world's most powerful neutron source in the world~\cite{ess}.
The neutrons will be produced by shooting protons to a heavy tungsten wheel in which the neutrons are produced by spallation.
A nearly \SI{400}{\meter} long linear accelerator will accelerate the protons to up to the required energy of \SI{2000}{\mega\electronvolt}.
In the linear accelerator, the protons are accelerated by precisely synchronized \gls{rf} fields.
Six different types of \gls{rf} cavities are used at \gls{ess} to apply the \gls{rf} field to the protons.
The \gls{llrf} system is responsible to control the amplitude and phase of the \gls{rf} field inside these cavities.

All low latency systems at \gls{ess}, which includes the \gls{llrf} system, will follow the \gls{mtca} standard~\cite{mtca}.
The \gls{mtca} standard defines a computing system with a specific size of boards, the communication between the boards over a backplane and the management of the system by \gls{ipmi}.
So-called \glspl{amc} are inserted into an \gls{mtca} crate which realize different functions and build up a specific computing system.
Generally, one of the \glspl{amc} is a \gls{cpu} board that performs the central calculations and establishes a link to the upstream systems.
Another common type of \gls{amc} is the \gls{fpga} board which allows application-specific functionalities to be implemented by them.
The MTCA.4 extension of the standard adds enhanced possibility for rearside I/O connections and precision timing~\cite{desy_mtca4}.
The rearside I/O connections are realized by plugging a separate board---the \gls{rtm}---from the backside into the crate that is connected to the \gls{amc} by a dedicated connector.

\begin{figure}[!b]
  \centering
  \includegraphics*[width=82.5mm]{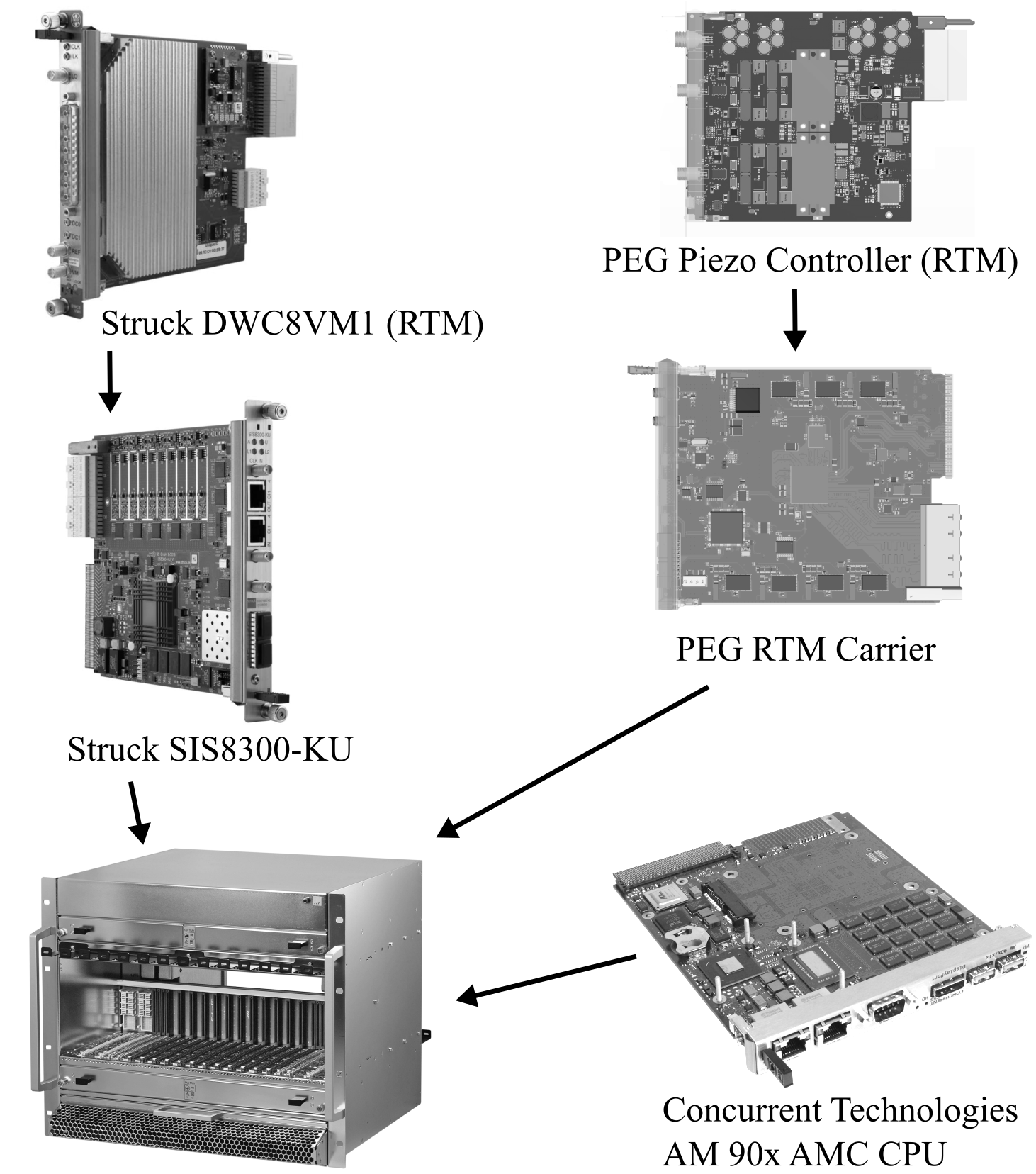}
  \caption{The \gls{mtca} boards building the \gls{ess} \gls{llrf} system.}
  \label{fig_boards}
\end{figure}

The \gls{llrf} controller of \gls{ess} is implemented on the \gls{fpga} of a Struck SIS8300-KU digitizer \gls{amc}~\cite{struck8300}.
Besides the control of the \gls{rf} field, the piezo controller used to tune the frequency of the cavity is also part of the \gls{llrf} system at \gls{ess}~\cite{ess_llrf}.
The piezo controller is realized on an \gls{rtm} developed by the \gls{peg}---a collaboration between Polish universities and research centers~\cite{peg}.
To control the piezo controller, an \gls{fpga} \gls{amc} (\gls{peg} \gls{rtm} carrier) has been developed by \gls{peg}.
Figure~\ref{fig_boards} shows all the boards of a typical \gls{llrf} system at \gls{ess} that go into an \gls{mtca} crate.
Other sub-systems of \gls{ess} also use the Struck SIS8300-KU.

All these applications implemented on an \gls{fpga} have to provide some common functionalities.
These functionalities are for example:
(1) Set parameters on the board and read the status from the board by the \gls{cpu} in the crate over \gls{pcie}.
(2) Acquisition of data and storage in a on-board memory.
(3) Configuration of the board peripherals.
For the same board, these functions are basically the same but also among different boards they require similar blocks of \gls{fpga} firmware.
An \gls{fpga} framework is providing these functionality and allows the re-use of code among projects.

Frameworks that are provided by the board suppliers can only be implemented on their boards. 
On the other hand, available frameworks developed in the research community base a lot on own implementations of central elements like buses~\cite{desy_framework}.
Therefore, the \gls{ess} \gls{fpga} Framework has been developed as a common platform for \gls{fpga} developments on different boards at \gls{ess}.
To enable extensibility, the framework incorporates existing standards whenever possible.


\section{Concepts of the Framework}

The \gls{ess} \gls{fpga} Framework simplifies the implementation of \gls{fpga} applications for \gls{mtca} systems with a \gls{cpu} in the crate.
To make it easy to adapt the framework to different \gls{fpga} boards, board-specific parts are separated from common parts and the interfaces between those are well-defined.
The same applies to the interface between the actual application and the framework.
All project-specific changes are controlled by central configuration files.

On the one hand, it is important for a facility like \gls{ess} to receive support for the systems for several years.
Therefore, solutions where the source code is not available and future support is uncertain should be avoided.
On the other hand, it is expensive to develop all code from scratch.
The tradeoff that the \gls{ess} \gls{fpga} Framework makes, is to follow open standards as far as possible.
Specifically, the base communication infrastructure is based on \gls{axi4}---and open-standard, on-chip interconnect developed by ARM~\cite{axi}.

\gls{axi4} is widely used and is also the standard interconnect to build up systems-on-chip in Xilinx Vivado~\cite{vivado_axi}.
This allows using Xilinx and third-party \gls{ip} blocks within the framework.
Many different functions are available in the \gls{ip} repository of Xilinx~\cite{xilinx_ip}.
Xilinx as the chip supplier has an interest in a living ecosystem around their \glspl{fpga} and, therefore, the availabilty of their \gls{ip} blocks.
Even if an own solution is preferred, such an \gls{ip} block can help to get a first prototype of the system.

A common framework for different projects simplifies the development of the software on the upstream \gls{cpu}.
This affects the kernel driver as well as the access by the control system---\gls{epics} in case of \gls{ess}~\cite{epics}.
The base of the software can be shared between the projects.
Board- and project-specific parts can be added as modules.

Finally, the \gls{fpga} framework should integrate well with the workflow and tools used for software development at \gls{ess}.
One example here is version control by git~\cite{git}.
Many of these concepts are not yet widely used for \gls{fpga} developments but suit it well.


\section{Structure of the Framework}

\begin{figure}[!htb]
  \centering
  \includegraphics*[width=82.5mm]{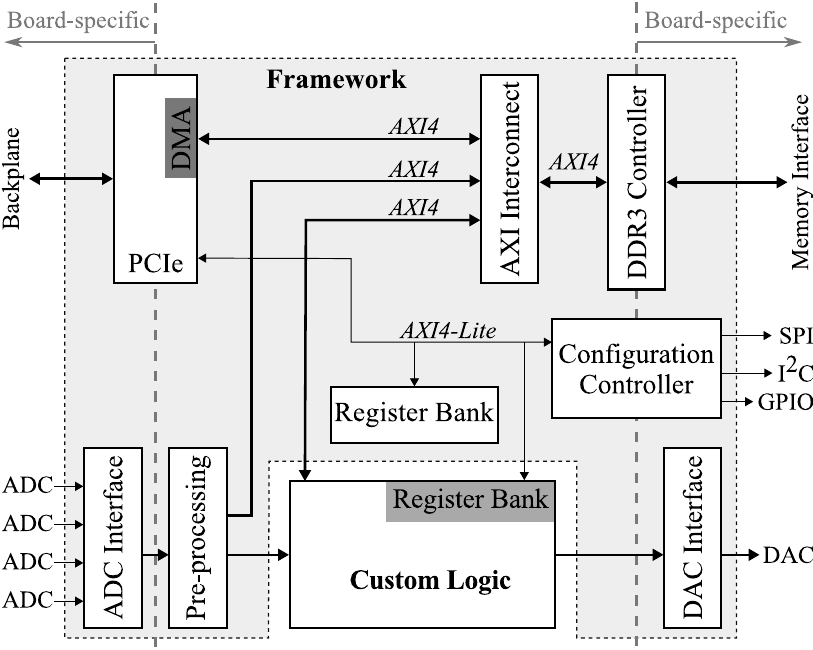}
  \caption{Structure of the ESS FPGA Framework.}
  \label{fig_overview}
\end{figure}

The \gls{ess} \gls{fpga} Framework consists of different modules, which are illustrated in Figure~\ref{fig_overview}.
The application-specific functionality is contained within the custom logic which interfaces with the actual parts of the framework: the core framework and the data path.
Whereas the core framework is the central part that controls the communication between the FPGA blocks and the external components as well as the upstream \gls{cpu}.
The components of the data path are responsible for the processing of high-speed and high-throughput data.

\subsection{The Core Framework}

Three functionalities are part of the core framework: the upstream communication to the \gls{cpu} in the crate, the management of the on-board memory and the configuration of the board peripherals as described in the next section.

Within \gls{mtca} crates, \gls{pcie} is generally used for the upstream communication with a \gls{cpu} board
The upstream communication can be split into two different types.
Two different interfaces are provided by the \gls{pcie} block of the \gls{ess} \gls{fpga} Framework for the two communication types.

Firstly, the \gls{cpu} needs access to the on-board memory in order to transfer data from and to the board.
An \gls{axi4} interface with an additional \gls{dma} engine is directly connected to the memory controller of the framework.
Secondly, parameters that control the functions on the board need to be set and the status of the board needs to be supervised.
A separate \gls{axi4}-Lite interface enables the access to one or multiple register banks within the framework.

An \gls{axi4} interconnect is used to combine and arbitrate the memory accesses from the different modules in the framework.
A memory controller connected to the \gls{axi4} interconnect realizes then the access to the on-board memory.

\subsection{Configuration of the Board Peripherals}

An essential functionality of the core framework is to configure the peripherals of the \gls{fpga} board.
These peripherals could be for instance \glspl{adc}, \glspl{dac}, a clock generator or the clock distribution.
Slow and simple serial protocols,\ such as \gls{spi} and \gls{i2c}, are generally used for the configuration of these peripherals.
Other components and functionalities are controlled by dedicated pins via a \gls{gpio} interface.

In many projects, the configuration of the peripherals is controlled by the device driver running on the \gls{cpu} in the crate.
So after booting the system, the \gls{cpu} in the system sends the configuration commands over registers on the board to dedicated interfaces in the \gls{fpga}.
These interfaces issue the according signals to the configuration buses of the peripherals.
As the used serial buses are not completely standardized, it is usually necessary to develop a specific firmware block to reflect the slighlty different behaviour of each peripheral.

The board configuration is also rather static for most applications, i.e.\ it is not changed or just to a small extent during operation.
Based on this static characteristic, the \gls{ess} \gls{fpga} Framework follows a different approach for the board configuration.
The idea is to bundle all functionality that belongs to the board.
The configuration of the peripherals goes together with the \gls{fpga} configuration rather than the board driver.
For this reason, a small microcontroller system based on a Xilinx MicroBlaze soft-core~\cite{vivado_microblaze} is included in the \gls{fpga} design that controls the configuration procedure.
As a very basic setup of a MicroBlaze system is sufficient for executing the configuration, the penalty in terms of \gls{fpga} resources is small.

The software for the microcontroller is combined with the \gls{fpga} configuration (bit-file) and both together build the firmware of the board.
This approach makes it easier to use the serial interface \gls{ip} blocks provided by Xilinx because varieties in the protocols are easier to implement in the software of an on-chip controller than on the \gls{cpu} of the crate.
It also simplifies the driver development as the configuration software is implemented by the \gls{fpga} engineers who are more familiar with the hardware-specific functions.
Nevertheless, the microcontroller is connected to the register bank and important board functions can be controlled by the \gls{cpu} and the status of the peripherals can be read.

\begin{figure}[!htb]
  \centering
  \includegraphics*[width=82.5mm]{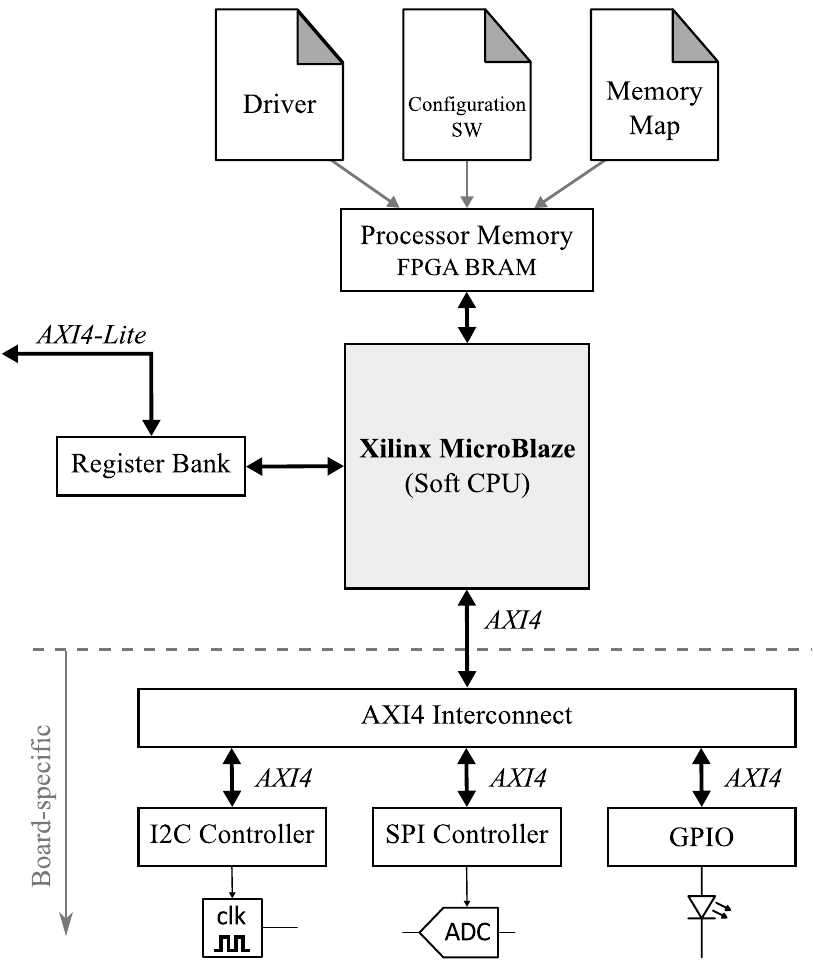}
  \caption{Block diagram of the Xilinx MicroBlaze system building the configuration controller.}
  \label{fig_config_system}
\end{figure}

\subsection{Data Path}

For high-speed and high-throughput data exists a separate data path.
The current implementation for the Struck SIS8300-KU digitizer uses the \glspl{adc} as the data source.
After the source, the data is passed to a pre-processing block which is not board-specific.
It is modular and may contain common functions such as filters, decimators and near-IQ sampling blocks.
A set of compatible blocks is part of the framework.
The pre-processing block contains also a memory interface through which raw or processed data can be stored directly in the on-board memory.
The pre-processed data is routed to the custom logic and is finally sent to a sink which is a \gls{dac} in the current \gls{llrf} application.

To allow the highest throughput and real-time, a simple protocol with data signals and a data valid signal is used for the data path.
To increase the consistency with the rest of the framework, the usage of the \gls{axi4}-Stream protocol is foreseen~\cite{axi_stream}.

\subsection{Integration of the Application---The Custom Logic}

The actual \gls{fpga} application is embedded into a wrapper with well-defined interfaces towards the \gls{ess} \gls{fpga} Framework.
Among other things, the wrapper defines interfaces for read and write access to the memory, inputs for data streams, and outputs for data streams.

\begin{figure*}[!htb]
    \centering
    \includegraphics*[width=\textwidth]{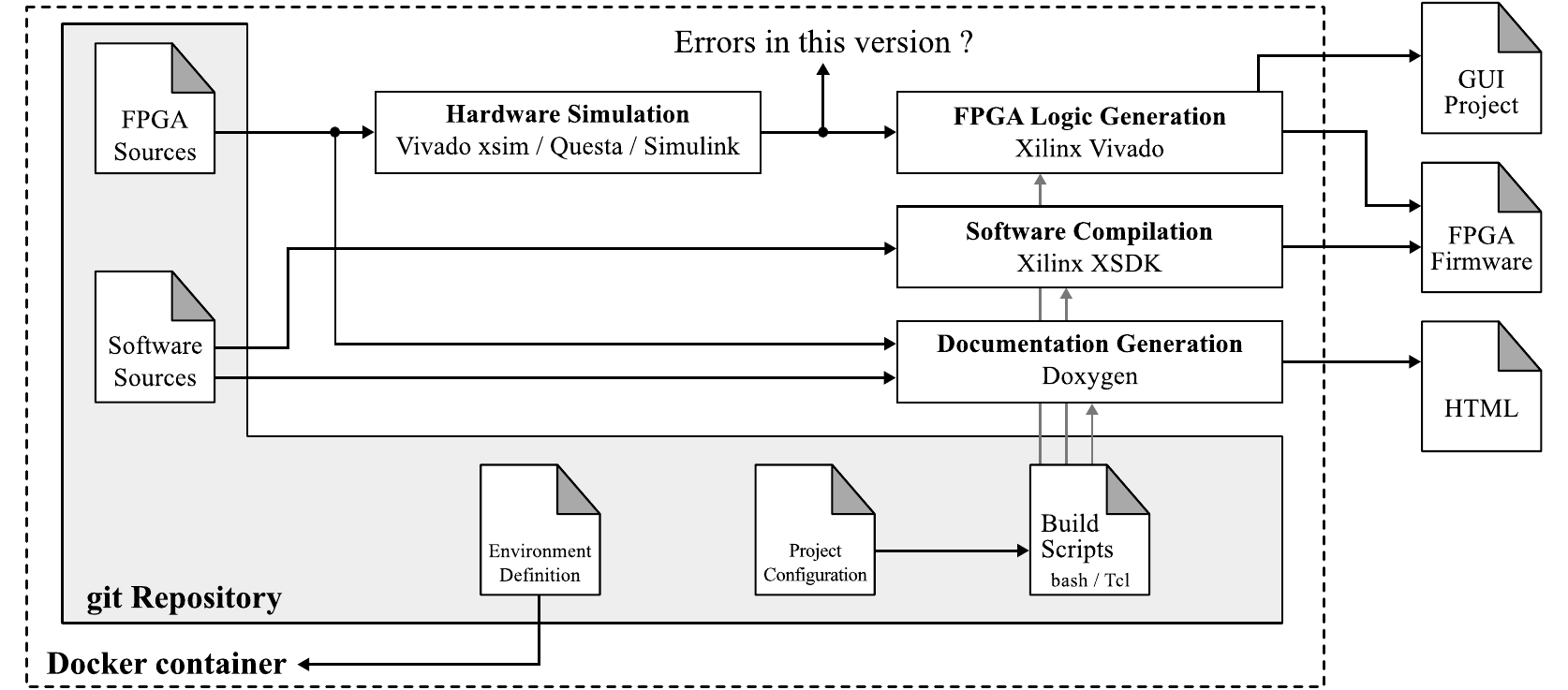}
    \caption{Steps and tools involved in the automated design process of the \gls{fpga} framework.}
    \label{fig_workflow}
\end{figure*}

Beside the defined interfaces, the custom logic also shares the register bank design with the framework.
The custom logic contains its own instance of the register bank with its own register map but is connected to the same \gls{axi4}-Lite bus as the register bank of the framework.
Thereby, the partitioning of the register bank is transparent to the driver.
It is even possible to split up the custom logic register bank to multiple banks for different sub-blocks.
The mapping of the different register banks to their corresponding addresses is handled by the design tools.


\section{Automated Design Process}

Along with the development of the \gls{ess} \gls{fpga} Framework, an automated workflow for \gls{fpga} projects has been developed.
The goal of this workflow is to achieve repeatability throughout the whole design process.
Therefore, developer A can take a version of a design and will get the same results that developer B got two weeks ago.
Another goal of the automation is that the continuous integration tool Jenkins~\cite{jenkins} can be used which is already part of the software integration at \gls{ess}.

\subsection{Tools in the Workflow}

The whole process from the source code to the final design that can be loaded to the \gls{fpga} involves many different tools.
Figure~\ref{fig_workflow} shows the complete workflow with the tools involved.

To orchestrate the different tools, GNU make is applied---a tool widely used for build automation in software development~\cite{make}.
GNU make is called with a certain goal that a developer wants to achieve.
From its configuration, GNU make knows which scripts need to be called to achieve this goal.
The scripts---written in languages like bash, Tcl and Python---control then the design tools.
These scripts are also part of the project code.

The first step in the automated design process is the testing of the code by running test benches on a module, sub-system and system level.
To allow the design process to be fully automated, the test benches need to be self-testing, i.e. no interaction by a designer is necessary to check the results.
Only if the design passes all tests successfully, the generation of the \gls{fpga} configuration and the software for the configuration controller is executed.
Tools supported for testing are, for instance, the Vivado simulator, Mentor Graphics Questa and Simulink.
These tests are also part of the project code.

For the generation of the \gls{fpga} configuration, Xilinx Vivado is run in batch mode~\cite{vivado_tcl}.
Vivado generates first the required \gls{ip} blocks and block designs from their configuration.
Afterwards, the scripts drive Vivado to execute all the steps from synthesis over place and route to the bit-file generation.
During this process, Vivado stores all relevant information into report files.
To allow a designer to analyze the system, a project is generated that can be opened in the Vivado \gls{gui}.

In a similar way, Vivado XSDK compiles the software for the configuration controller~\cite{vivado_xsdk}.
Finally, both the bit-file and the software are bound to a single firmware file that can be loaded to the \gls{fpga}.

In parallel, doxygen is used to extract documentation about the modules and generate \gls{html} files out of them~\cite{doxygen}.
This is done for \gls{hdl} code as well as software.
To include \gls{ip} blocks and block designs in the documentation it is recommended to put the documentation into a \gls{hdl} wrapper around the \gls{ip} block.

Finally, to define the development environment the support for Docker containers will be added to the workflow.
Docker can be understood as a lightweight version of a virtual machine.
It defines an operating environment in which the software runs but does not need to start a complete virtual machine that occupies a lot of computer resources.
For the automated design process, a Docker container could define for example the tool versions and environment variables.

\subsection{Project Organisation}

The code base of a project following the described design process is managed by the version control system Git~\cite{git}.
As described before, the scripts which define how the output products of the project are generated are also part of the code.
Therefore, the automated design process is also under version control and part of the project.
The same applies to the test benches and test scripts.
By including docker containers, even the development environment. 
The result is that everything that defines the project is under version control.
This means that if a developer receives a version of the project, everything is included and the results can be reproduced as intended.

To achieve effective version control, a clean folder structure is necessary.
Therefore, all files---sources, configuration files, scripts---are organized in a folder structure that is defined by the automated design process and does not follow the folder structure imposed by the design tools.
A separate folder contains all the external modules and libraries that are used by the project.
The number of files that belong to the project is kept to a minimum and all files that can be generated from other files should not be part of the code base.
This is a general suggestion to make version control of the code work effectively.
For instance, only the configuration files (.xci) of the Xilinx \gls{ip} blocks are included in the project code as all the other files are re-generated by Xilinx Vivado.
Most of the generated files such as reports from Xilinx Vivado, the software development workspace of Vivado XSDK or the Doxygen module documentation are placed in dedicated folders to clearly separate them from the files that go into version control.


\section{Results and Resource Usage}

A first version of the \gls{ess} \gls{fpga} framework with support for the Struck SIS8300-KU digitizer has been developed.
The framework has been tested at a frequency of \SI{125}{\mega\hertz} and \SI{200}{\mega\hertz} for the central \gls{axi4} interconnect.

As Table~\ref{tab_resource_framework} shows, the resource usage is reasonable with most values around or below \SI{20}{\percent}.
Only the \gls{bram} usage going up to \SI{28.8}{\percent}.
With the applied configuration, debugging of the configuration controller software is possible on the board.
By removing this possibility and optimizing the code for the configuration processor, the \gls{bram} usage can decreased essentially.

\begin{table}[hbt]
   \centering
   \caption{Resource usage by the \gls{ess} \gls{fpga} Framework.
            FPGA: Xilinx Kintex Ultrascale xcku040-ffa1156-c, \gls{axi4} data width: \SI{512}{\bit}, tool: Xilinx Vivado 2017.1.}
   \begin{tabular}{lrrr}
       \toprule
       \textbf{Resource} & \textbf{Utilization} & \textbf{Available} & \textbf{Util. \si{\percent} } \\
       \midrule
           LUT           & \num{51463}          & \num{242400}       & \SI{21.2}{\percent} \\
           Flip-Flop     & \num{69973}          & \num{484800}       & \SI{14.4}{\percent} \\
           BRAM          & \num{173}            & \num{600}          & \SI{28.8}{\percent} \\
           DSP blocks    & \num{86}             & \num{1920}         & \SI{4.5}{\percent}  \\
           MMCM          & \num{1}              & \num{10}           & \SI{10.0}{\percent} \\
           PLL           & \num{3}              & \num{20}           & \SI{15.0}{\percent} \\
       \bottomrule
   \end{tabular}
   \label{tab_resource_framework}
\end{table}

%


\section{Conclusion}

With the \gls{ess} \gls{fpga} Framework, a platform has been created that simplifies the \gls{fpga} developments for \gls{mtca} boards.
Functionality that is needed for different projects is provided by this framework.
As the framework interconnect is based on the \gls{axi4} standard, \gls{ip} blocks from other developers can be integrated and development time may be saved.
The automated design process of the framework allows the \gls{fpga} designers to achieve repeatability of their results from testing over synthesis to integration.
After the successful integration of the \gls{llrf} application, beam diagnostic applications at \gls{ess} are currently ported to the framework.


\section{Acknowledgment}
We would like to thank Struck Innovative Systeme GmbH for their support in the
development of the ESS FPGA Framework.



\null

\end{document}